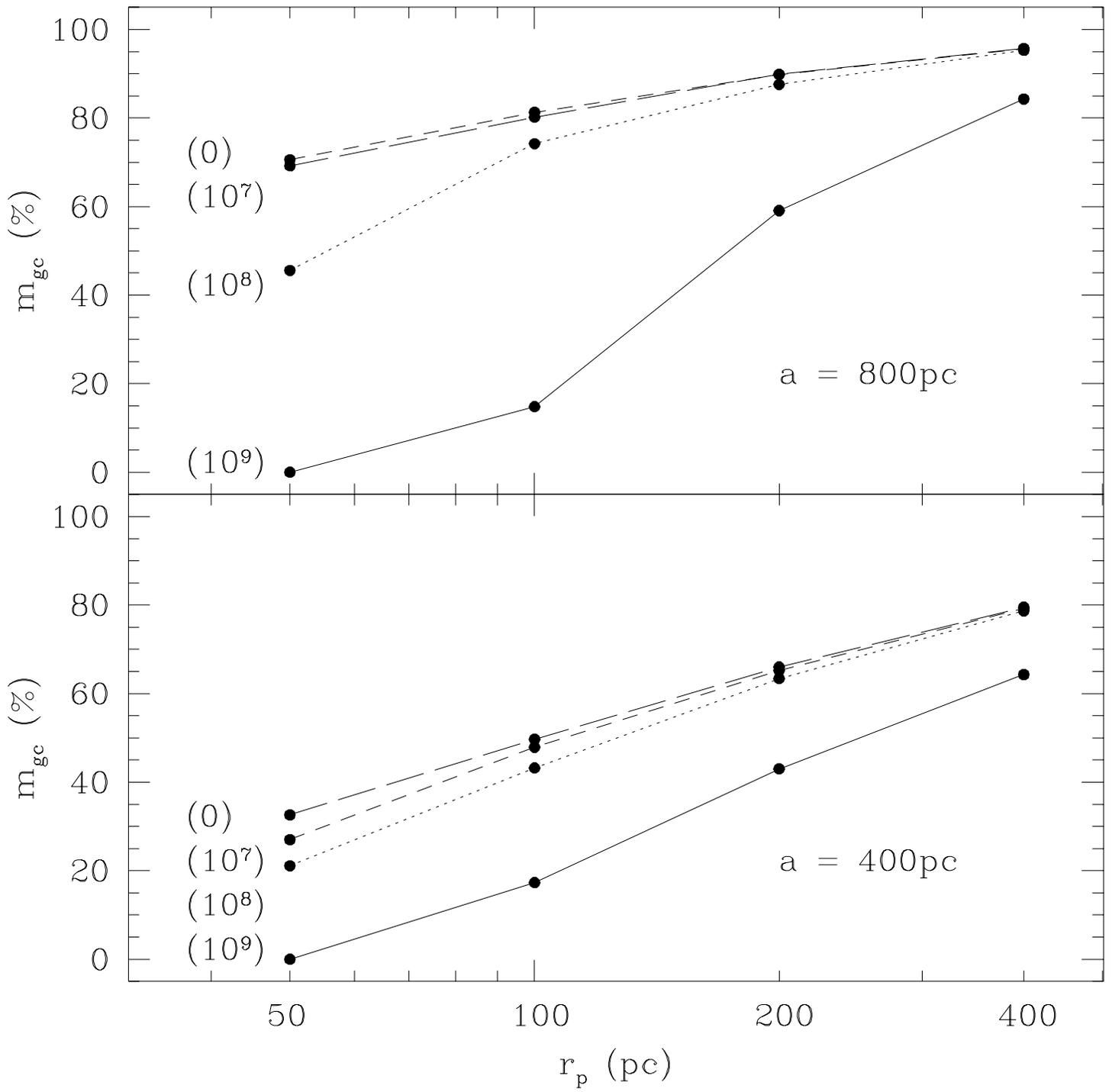

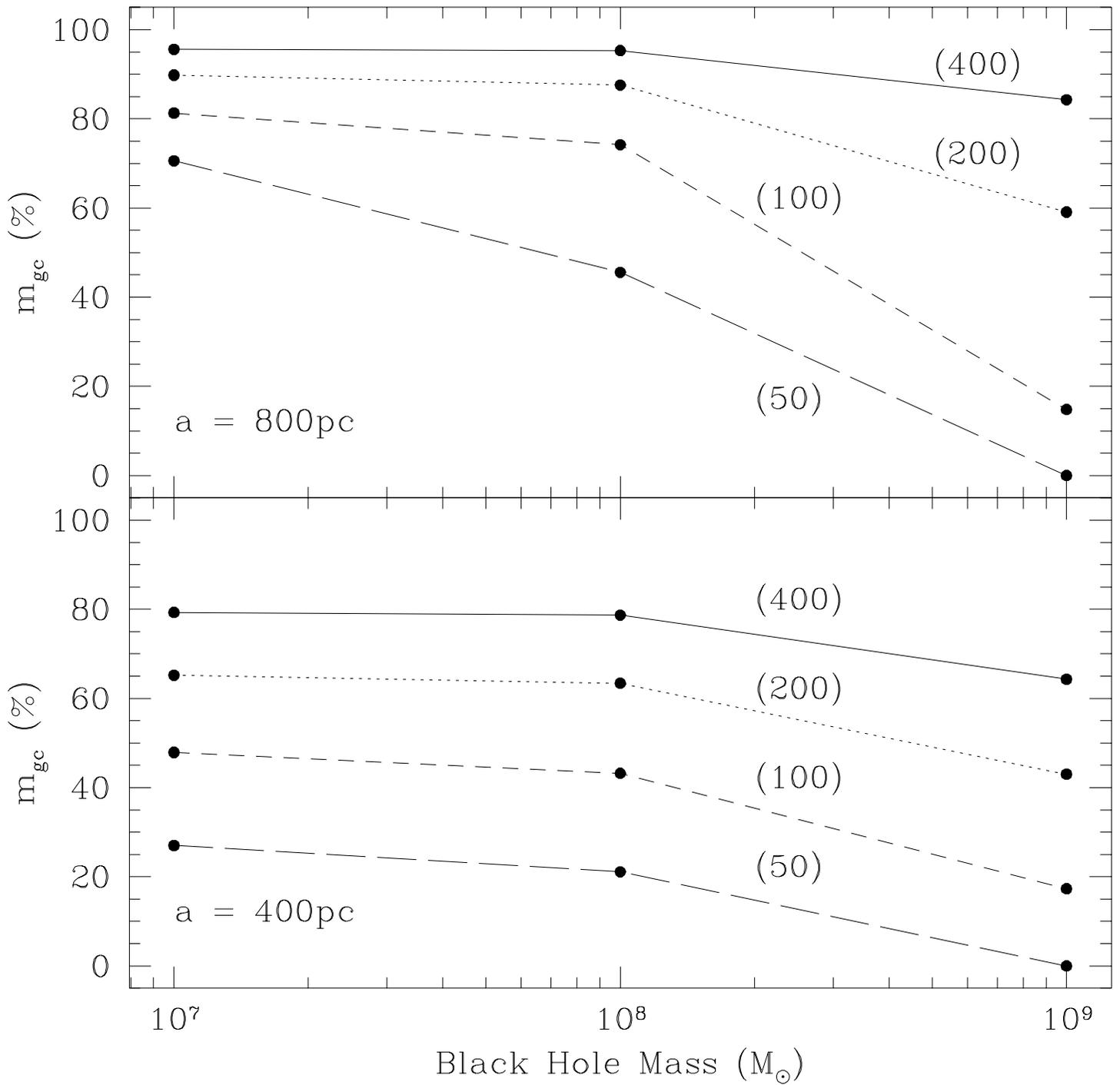

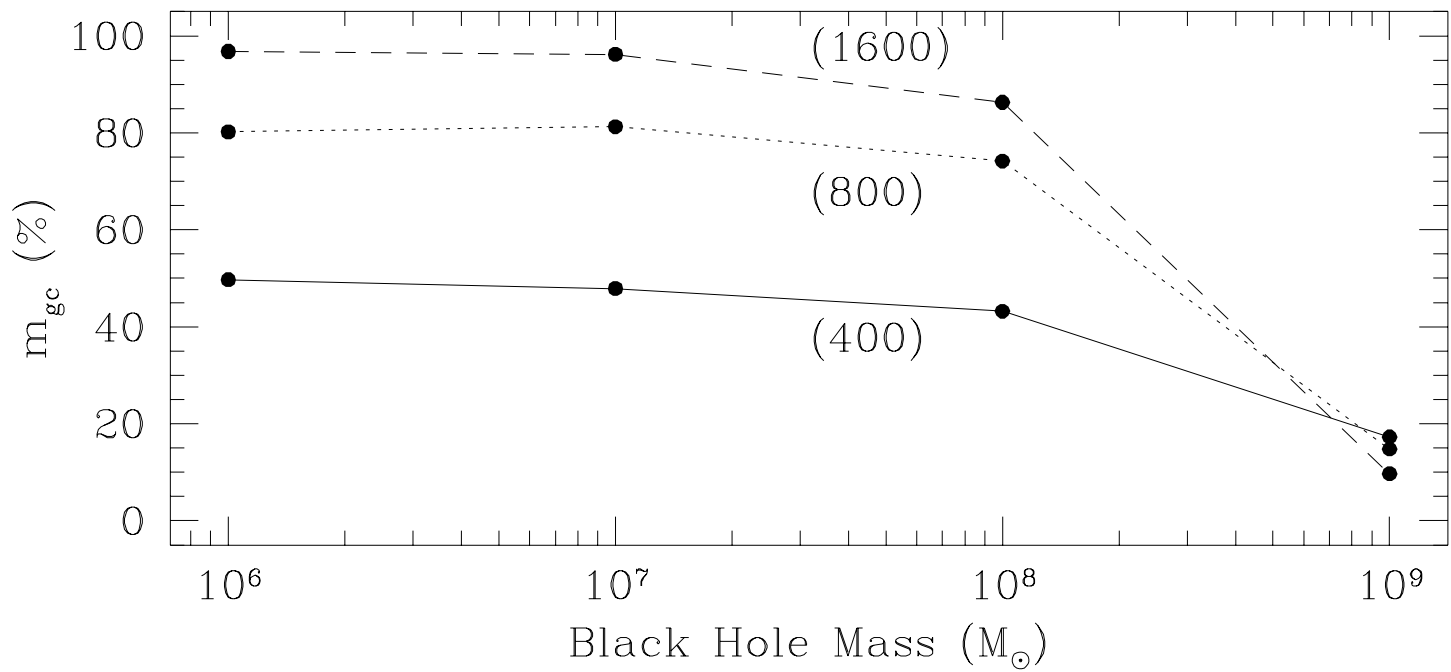

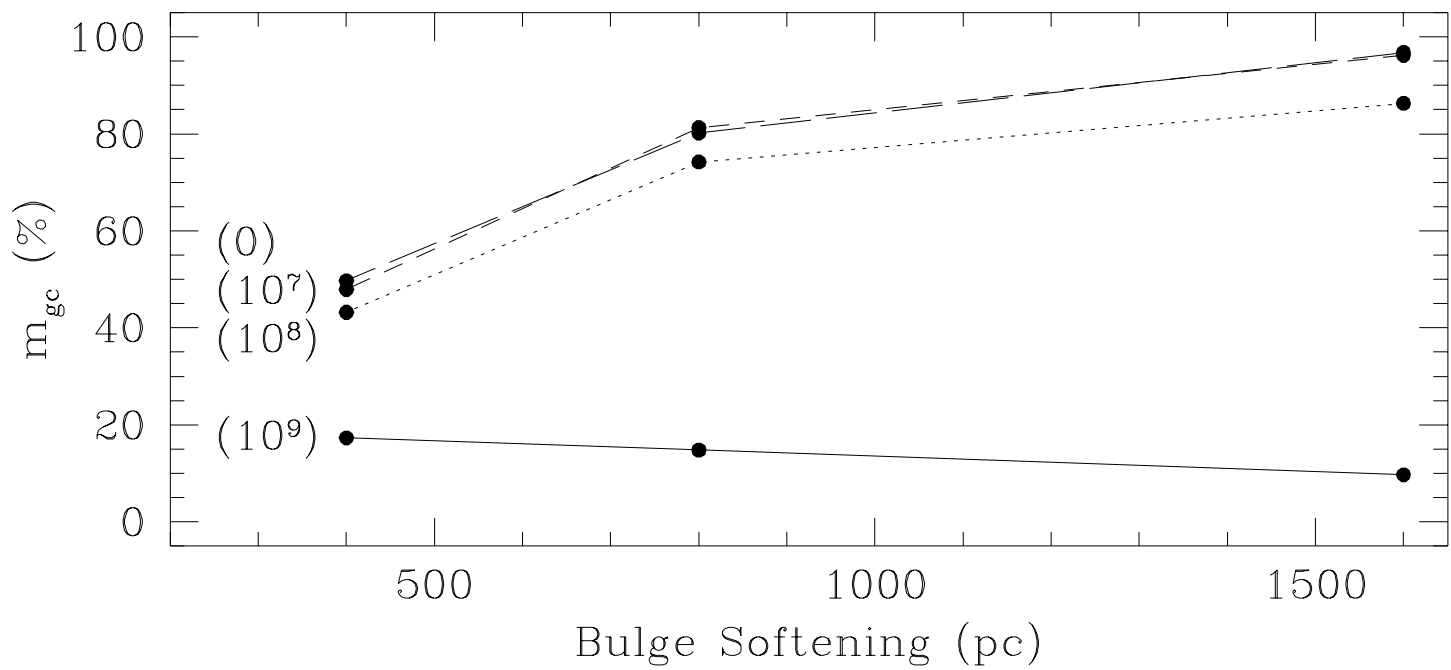

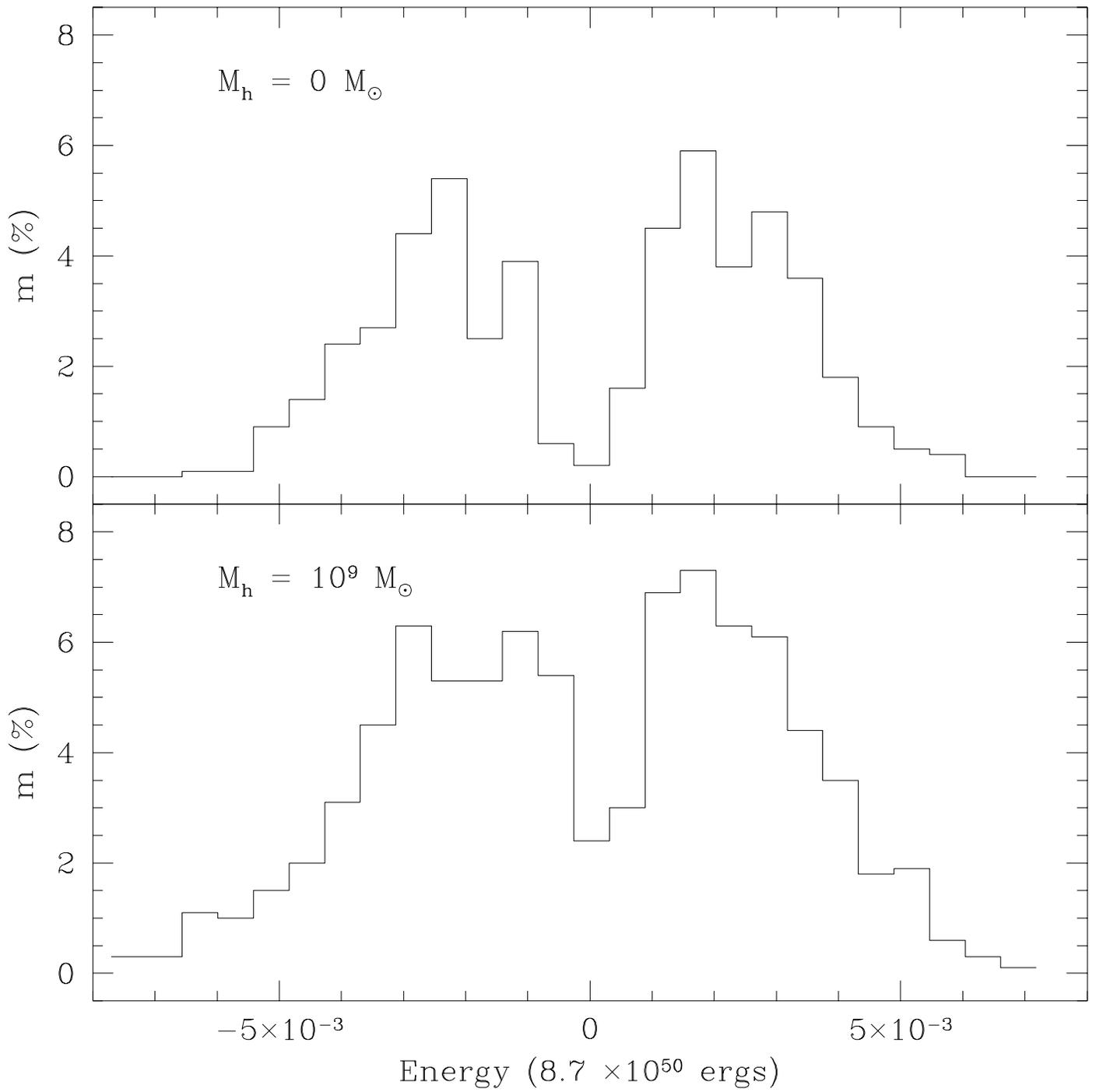

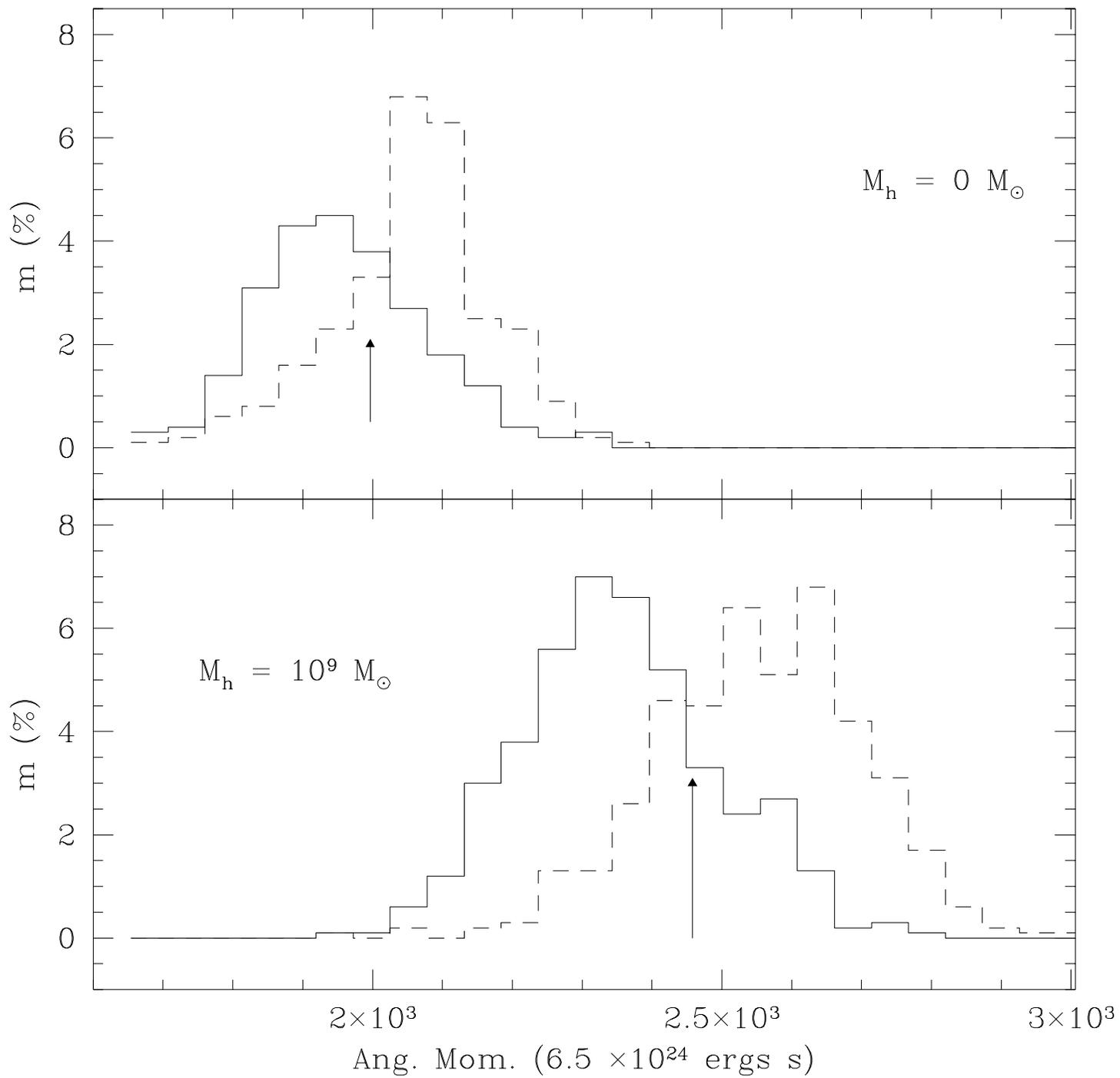

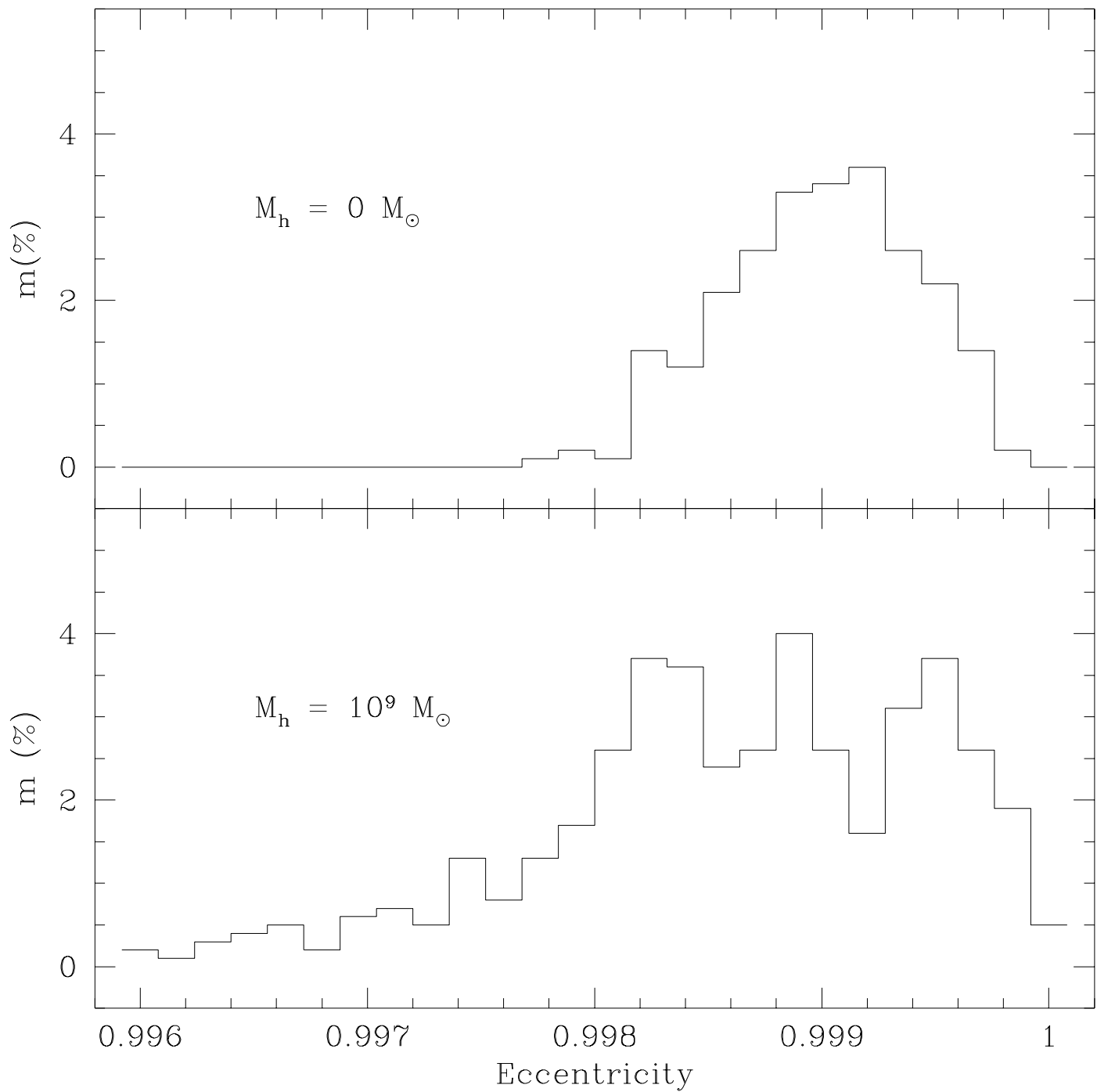

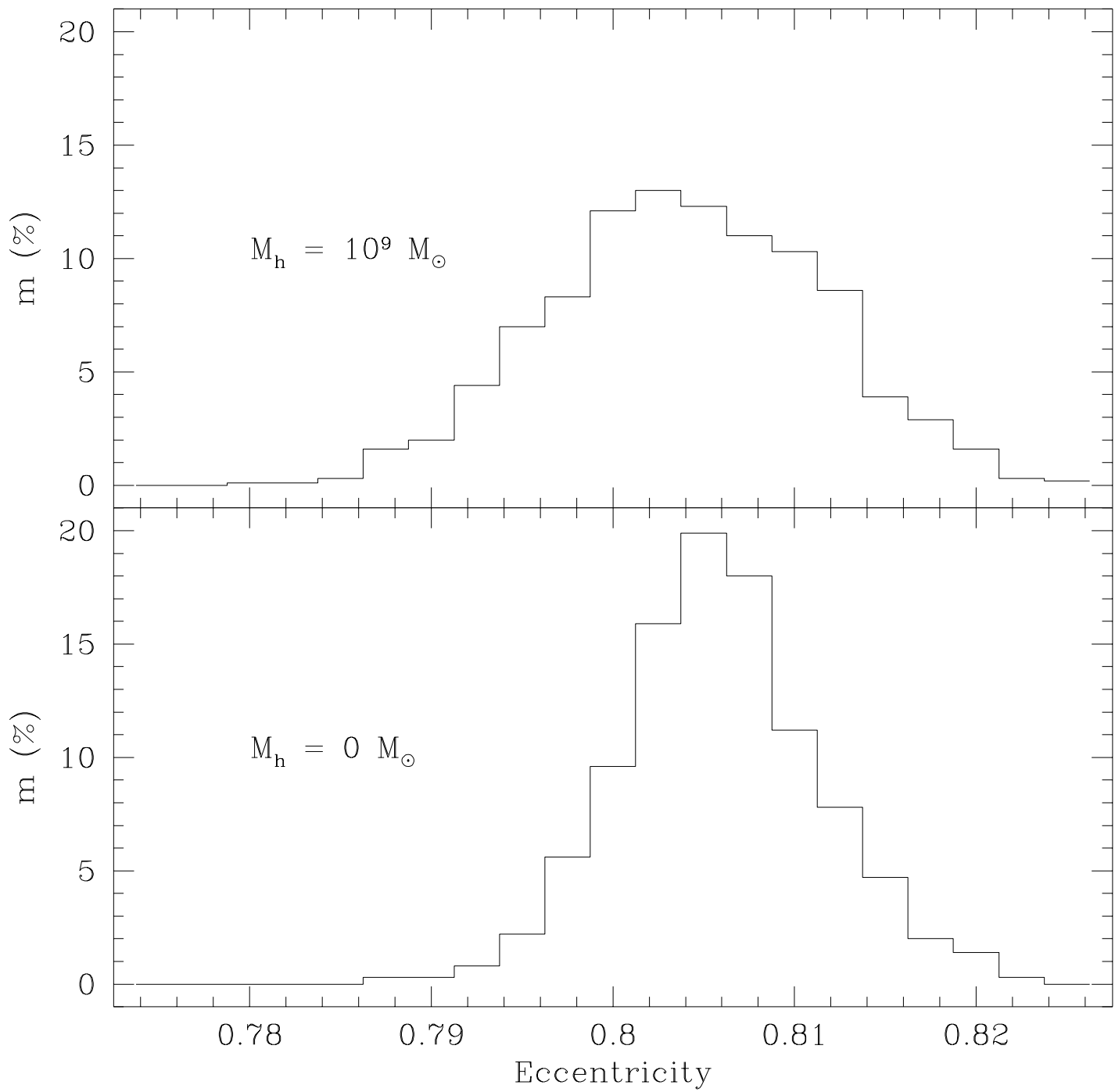

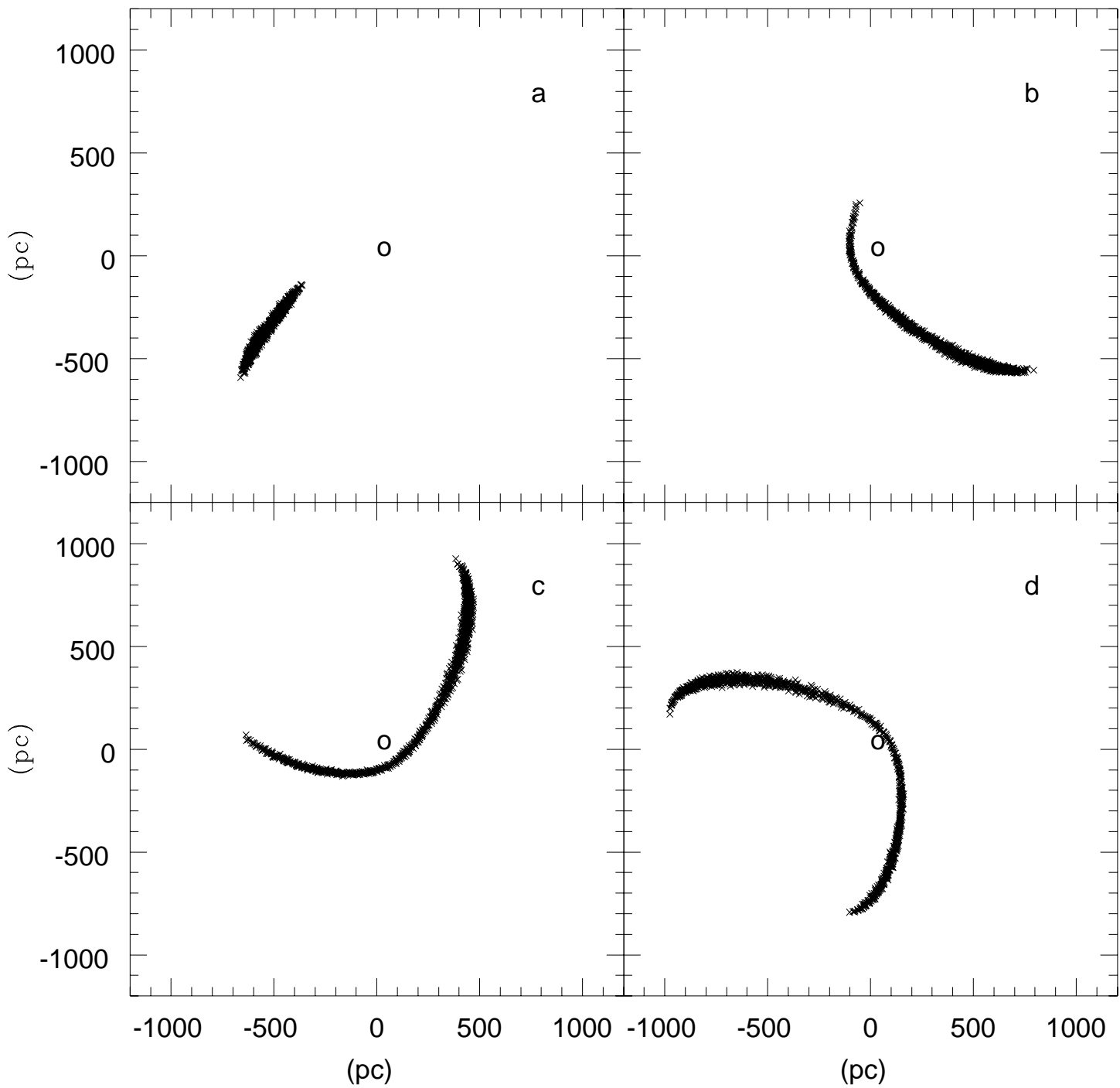

# Competition of Supermassive Black Holes and Galactic Spheroids in the Destruction of Globular Clusters

Jane C. Charlton and Pablo Laguna

Department of Astronomy and Astrophysics

and

Center for Gravitational Physics and Geometry

The Pennsylvania State University, University Park, PA 16802

## ABSTRACT

The globular clusters that we observe in galaxies may be only a fraction of the initial population. Among the evolutionary influences on the population is the destruction of globular clusters by tidal forces as the cluster moves through the field of influence of a disk, a bulge, and/or a putative nuclear component (black hole). We have conducted a series of N-body simulations of globular clusters on bound and marginally bound orbits through potentials that include black hole and spheroidal components. The degree of concentration of the spheroidal component can have a considerable impact on the extent to which a globular cluster is disrupted. If half the mass of a $10^{10} M_\odot$ spheroid is concentrated within 800pc, then only black holes with masses $> 10^9 M_\odot$ can have a significant tidal influence over that already exerted by the bulge. However, if the matter in the spheroidal component is not so strongly concentrated toward the center of the galaxy, a more modest central black hole (down to $10^8 M_\odot$) could have a dominant influence on the globular cluster distribution, particularly if many of the clusters were initially on highly radial orbits. Our simulations show that the stars that are stripped from a globular cluster follow orbits with roughly the same eccentricity as the initial cluster orbit, spreading out along the orbit like a "string of pearls". Since only clusters on close to radial orbits will suffer substantial disruption, the population of stripped stars will be on orbits of high eccentricity.

*Subject headings:* clusters: globular — galaxies: structure

## 1. Introduction

The present distribution of globular clusters in galaxies may differ substantially from the initial distribution. In the Milky Way, this is evidenced by the difference between the isotropic velocity distribution of globular clusters and the more radial distribution of other members of the spheroidal component (such as the RR Lyrae stars). In at least several cD ellipticals (Lauer & Kormendy 1986; Harris et al. 1991; McLaughlin, Harris, & Hanes 1993), including M87, the globular cluster distribution is less centrally peaked than the light. This could be due to initial conditions, but it seems likely that at least to some extent an evolution of the globular cluster distribution is involved (McLaughlin, Harris, & Hanes 1994).

Several mechanisms affect the destruction of globular clusters, and which is dominant depends on the type of galaxy and its stage of evolution. These mechanisms include cluster evaporation, dynamical friction due to passage through the galaxy, and shocking due to rapid changes in the gravitational potential as the cluster passes through the spheroid or disk. Bulge shocking was found to be the primary mechanism for the initial destruction of clusters in our Galaxy, affecting the region within 2kpc (Aguilar, Hut, & Ostriker 1988).

There is mounting evidence that supermassive black holes are present in the centers of galaxies, ranging from a $3 \times 10^9 M_\odot$ black hole in M87, to a possible $10^7 M_\odot$ one in M31, and perhaps even a small ($10^6 M_\odot$) one in the Milky Way. There is some additional indirect evidence that supermassive black holes exist in the centers of many galaxies. The popular theory for the source of power in active galaxies and quasars is gas feeding into a central black hole, which could have a mass of $10^8 M_\odot$ or more. Since the number of active galaxies and quasars was very large in the past, it seems that most normal galaxies may have once been in that state. The difference at present could merely be the lack of an adequate source of fuel.

If black holes are commonly found in the centers of galaxies, then they could also have an effect on the system of globular clusters. Ostriker, Binney, and Saha (1989) have found that in the case of M87, a $10^9 M_\odot$ black hole could affect the distribution of globular clusters out to radii of 8kpc, and would be particularly important if box orbits were dominant in the initial distribution. In general, although the central black hole is likely to be less massive than the spheroid, the spheroid is also less concentrated. Thus when a globular cluster passes through a spheroid that contains a supermassive black hole there must be some radius within which the black hole will have a dominant effect on the destruction.

It is the goal of this paper to explore the relative and combined effects of a spheroid and black hole system. N-body simulations have been performed with a modified tree-code, modeling the globular cluster using particles, and the spheroid and black hole as added analytic potentials. The disruption of the globular cluster has been studied as a function of the pericentric distance in a parabolic encounter, for various possible black hole masses. For comparison, an elliptical orbit was also considered. We computed the orbital parameters of the stars that were stripped from the globular cluster.

In §2 we describe the initial conditions for globular clusters, spheroids, black holes, and orbital parameters, as well as the numerical method used for conducting our simulations. In §3 we present the results of our simulations for the various cases. The fractions of globular cluster mass that have various fates is given as a function of the distance of closest approach to galactic center for various black hole masses and degrees of bulge concentration. The orbital properties of particles that are stripped from the cluster are also presented. Finally, in §4 we discuss the circumstances in which the effects of supermassive black holes, in conjunction with spheroids, can be important for destroying globular clusters. Relevance of our general study to various types of galaxies is also addressed.

## 2. Initial Conditions and Numerical Method

Simulations consist of globular clusters (represented by 1000 particles) following bound (elliptic) or marginally bound (parabolic) orbits. The initial position of the globular cluster in each simulation was chosen well outside the half-mass radius of the spheroid. For marginally bound orbits, the initial velocity of the center of mass of the cluster is determined once the pericentric distance is specified. The simulations are conducted using a modified N-body tree code. That is, the forces among the particles representing the globular cluster are computed via a tree algorithm; on the other hand, the forces on each particle due to the spheroid/black hole system are calculated from analytic potentials.

The amount of material that is stripped in a passage through the bulge depends upon the distribution of stars within a globular cluster. There is tremendous variety in the properties of globular clusters, and specifically in how centrally concentrated they are. We represent a globular cluster by a King model with a central potential $W_0 = 4$, and a half-mass radius of 10pc. The mass of the cluster $M_{gc}$ is our basic mass unit, and the black hole and spheroid masses will be expressed in these units. All of the 1000 particles modeling the globular cluster are taken to have equal mass. The cluster half-mass radius that we have chosen is within the observed range for Galactic globulars, but is on the high end. It is particularly unusual to have a value this large in the central 10kpc of our Galaxy (van den Berg 1994), probably because these are the clusters that are most easily destroyed. Such clusters are, however, likely to have been more common in the initial distribution.

For the spheroid, we have chosen a simple potential that could represent the bulge of a spiral galaxy or the nuclear region of an elliptical. The potential has the form (Hernquist 1990)

$$\phi(r) = -\frac{GM_{bg}}{r + a} \tag{1}$$

which results from a mass distribution of the form

$$M(r) = M_{bg}\frac{r^2}{(r + a)^2} \tag{2}$$

The half mass radius of the spheroid is then $(1 + \sqrt{2})a$. The spheroid mass is set to $10^5 M_{gc}$ in all of the runs, but we will discuss later the effect of changing this value. We consider values ranging from $a = 400$pc to $a = 1600$pc for the "softening" parameter of the bulge.

A range of black hole masses were used, as well as a case with a spheroid only and no black hole. Generally, runs were conducted with $M_{bh} = 0, 10^2, 10^3$, and $10^4 M_{gc}$. All runs are listed in Table 1, along with the results for the amount of disruption that are described in the following section.

For simplicity, we mostly considered initially parabolic orbits. (For comparison results for an $e = 0.8$ orbit are also examined.) An initial distribution biased toward radial orbits is expected to arise in a galaxy with a tri-axial potential (Long, Ostriker, & Aguilar 1992). Simulations are started outside the region of the spheroid within which 95% of the mass lies. (Results are found to be insensitive to this choice.) The effect of varying the pericentric distance, $r_p$, over the range 50pc to 400pc, is studied for two series of cases with $a = 400$pc and $a = 800$pc. For $r_p$ fixed at 100pc, runs with 1600pc were conducted.

The tree code requires the choice of a softening parameter $\epsilon$ and of a fixed time-step $\delta t$. The softening parameter is set small enough so that the cluster does not evaporate by relaxation due to over-softening. The time-step must be set small enough to follow close interactions accurately, without being too demanding on computing resources. Experimentation yielded $\epsilon = 0.05$pc and $\delta t = 0.05$ time units as the best choices. (For convenience, the code is run with 1 pc as the unit of distance, and with the globular cluster mass as the unit of mass for $G = 1$. In this system of units, for a globular cluster with mass $10^5 M_\odot$, a velocity unit corresponds to 20.8 km/s and a time unit to $4.6 \times 10^4$yrs.)

Several timescales are relevant to this problem. In particular the cluster will be affected by tidal stripping only if the relaxation timescale is longer than the time of the encounter. The crossing time for a cluster with mass $M_{gc} = 10^5 M_\odot$ and half mass radius $R_{hm} = 10$pc is:

$$t_{cr} = \sqrt{\frac{2\,R_{hm}^3}{G\,M_{gc}}} \sim 2 \times 10^6 \text{yrs}. \tag{3}$$

The relaxation time within the cluster depends on the number of particles, $N$, used in the simulation and can be estimated to be (Chernoff & Weinberg 1990):

$$t_{rlx} = \sqrt{\frac{M_{gc}R_{hm}^3}{2G}}\frac{0.138N}{M_{gc}\ln N} \sim 2 \times 10^7 \text{yrs}. \tag{4}$$

The timescale for the encounter can be estimated as:

$$t_{enc} = \sqrt{\frac{r_t^3}{G\,M(r_t)}} \tag{5}$$

where $r_t$ is the tidal radius and $M(r_t)$ is the mass within this radius. Even for the most extreme case with of a bulge concentration $a = 400$pc and a black hole mass $M_{bh} = 10^4 M_{gc}$ eq. (5) yields $t_{enc} \sim 10^6$yrs. This is more than an order of magnitude less than $t_{rlx}$ (eq. (4)), so globular clusters represented by 1000 particles can be disrupted by tidal forces.

## 3. Destruction of Globular Clusters

For each of the N-body simulations, Table 1 gives in its final three columns: $m_{gc}$, the percentage of the mass that remains bound to the globular cluster, $m_b$, the percentage of the stripped mass that is still bound to the bulge/black hole system, and $m_u$, the percentage of the stripped mass that is formally unbound from the bulge/black hole system. The fraction of mass that is stripped by tidal forces is computed after the cluster has passed the center of the galaxy and is well outside the half mass radius again.

Generally, the cases with a smaller $a$ (bulge softening) have more stripped mass because the tidal fluctuations are more extreme. This is illustrated in Figs. 1a and 1b which give, for $a = 400$pc and 800pc respectively, the fraction of mass that remains bound to the globular cluster as a function of the pericentric distance, $r_p$. The different curves represent cases with no black hole, and with black hole masses of $10^2$, $10^3$, and $10^4 \mathrm{M}_{gc}$. For the purposes of illustration, the globular cluster has been assigned a mass of $10^5 \mathrm{M}_\odot$, but the simulation results can be scaled to any value. Not only is it clear that the case with lower $a$ has stripped considerably more material from the globular cluster, but it is also evident that the black hole has only a small additional effect (above the spheroid alone) in most cases. Only in the extreme situation of a $10^9 \mathrm{M}_\odot$ black hole ($10^4 \mathrm{M}_{gc}$), does the black hole significantly affect the results, and the effect diminishes with increasing $r_p$. This is further illustrated in Figs. 2a and 2b which display the same data, showing the effect of changing the black hole mass for curves that represent orbits with different $r_p$ (the largest value resulting in the least mass stripped). The curves are closer to horizontal for $a = 400$pc (Fig. 2b), except for the smallest $r_p$. For the larger value of $a$, a black hole as small as $10^8 \mathrm{M}_\odot$ is seen to strip an additional $\sim 20\%$ of the cluster mass for $r_p = 50$pc.

Figure 3a illustrates the result of changing $a$, with $r_p$ fixed at 100pc, for various black hole masses. Here, results have been obtained for the additional case with $a = 1600$pc, in order to probe the regime of phase space where destruction by the spheroid is not important. If a black hole is absent (represented by $10^6 \mathrm{M}_\odot$ on the horizontal scale), 20% of the globular cluster mass is stripped for $r_p = 100$pc in the case with $a = 800$pc. The spheroid with a larger $a$ (1600pc) has a negligible shocking effect. This figure also illustrates the point that when the spheroid is relatively concentrated, it is the change in its parameters that affects the disruption, and the black hole mass has little effect. However, when $a$ is large the black hole mass is the most important factor in determining how much stripping will occur. Fig. 3b, giving the dependence, on $a$, of the fraction of the globular cluster that is disrupted for various black hole masses, better illustrates the minimal effect of a black hole with mass less than $10^8 \mathrm{M}_\odot$.

Even when $a$ is large and the black hole is large enough to be the dominant disruptive influence, the spheroid can still have an indirect effect on the disruption. The amount of spheroid mass inside the tidal radius determined by the black hole will determine the length of time spent there. This can be illustrated by comparing the cases with various $a$ for $10^9 \mathrm{M}_\odot$ black holes. We might expect that the larger direct influence of stripping in the more concentrated spheroid would lead to a larger mass loss in that case. However, Fig. 3a shows that the less concentrated bulge leads to a larger effect (i.e. the curves cross). This is due to the greater length of time that is spent near the galaxy's center.

The stars that are stripped from a globular cluster can either remain bound to the spheroid/black hole system or they can become unbound from the system. The fractions of the original cluster mass in these two components, $m_b$ (stripped-bound) and $m_u$ (stripped-unbound), after the cluster has moved well outside the half-mass radius are given in columns 5 and 6 of Table 1. As in the case of tidal stripping from individual stars by a close approach to a black hole (Laguna et al. 1993), about half the stripped mass remains bound and the other half is unbound. We should note that this distinction between bound and unbound material is somewhat artificial. The unbound material will still be bound to other components of the galaxy and will be likely to remain in an orbit not dissimilar from one bound to the bulge/black hole system.

The energy distribution for the stripped material is shown in Fig. 4a for the runs with $a = 400$pc and two extreme cases of black hole masses, 0 and $10^9 M_\odot$. This figure shows the separation of the debris into material with positive and negative energy, i.e. bound and unbound to the spheroid/BH system. The distribution of energies shows a wider spread in the case with a more massive black hole. In the case with $a = 1600$pc there is an even larger difference between the spread in energy of debris between cases with and without black holes. We have chosen to illustrate the $a = 400$pc case because this is the case for which the overall effect is largest due to contributions from both spheroid and black hole.

For the same two cases, with black hole mass 0 and $10^9 M_\odot$ and with $a = 400$pc, the angular momentum distribution is shown in Fig. 4b, separately for the positive and negative energy debris. The bound debris (solid histogram) shows a smaller median value of angular momentum than the unbound debris (dotted histogram). This is expected since, in order to satisfy the energy criterion for being unbound, this material should have a larger velocity. The difference in the overall median value (the initial angular momentum of the center of mass of the globular cluster) is simply due to setting the initial velocity such that the globular cluster passes 100pc from the center. However, the spread in angular momentum is larger in the case with the most massive black hole, similar to the effect in the energy distribution.

The increased spreads in energy and angular momentum distributions of debris, as tidal forces become more substantial, would be expected to produce a spread in the eccentricity distribution of bound material. This can be seen to be the case in Fig. 4c, that shows the eccentricities of particles that remain bound to the spheroid/BH system. These are estimated when these particles are far from the center of the galaxy, using the approximation that the mass enclosed is at a point at the center of the system. The eccentricity distribution is found to be wider in the case with a super-massive black hole, however, the difference is negligible. If the globular cluster is initially on a parabolic orbit, the orbits of the stars that are stripped from it will be very close to parabolic.

An additional two runs were conducted to test if there is a general rule that the debris follows the same orbit as the globular cluster. An eccentricity of 0.8 was chosen for this test. A smaller value would have required us to begin the orbit well within the spheroid mass distribution. The other parameters for the system were $a = 400$pc and $r_p = 100$pc, and two runs have black hole masses of 0 and $10^9 M_\odot$. These cases have similar amounts of material stripped to the analogous cases with $e = 1$. The eccentricity distributions are shown in Fig. 5. Again, there is larger spread in the case with a larger black hole, but the debris still has nearly the same eccentricity as the cluster from which it was stripped.

The material that is stripped from a globular cluster follows a similar path to its parent, but the small spreading in properties reflects itself in an interesting phenomenon. A time sequence is shown in Fig. 6, depicting the debris as it moves through the spheroid. The gradual spreading along a line is reminiscent of the "String of Pearls" effect noted in Comet Shoemaker-Levy during the time of its approach to Jupiter.

## 4. Physical Interpretation and Consequences

This study has allowed us to analyze the relative importance of two of the destruction mechanisms that will be be relevant to the evolution of globular cluster systems in galaxies. For a concentrated spheroid (with $a < 800$pc) the spheroid properties are the most important factors that influence the amount of mass stripped from a globular cluster. The presence or absence of a supermassive black hole makes little difference except in very extreme cases of mass or very close approaches. For less concentrated spheroids, the supermassive black hole can be responsible for the majority of stripping that occurs. If there is a $> 10^8 M_\odot$ black hole in a spheroid with $a = 1600$pc, 14% of a globular cluster's mass can be stripped if it passes within 100pc. Since a cluster may complete numerous orbits over the history of the galaxy an even wider approach could have a substantial effect integrated over time.

Many of the results we have obtained can be understood qualitatively with the rough analytic expression for the tidal radius. For a spheroid mass $M_{bg}$ and central black hole mass $M_{bh}$ with potential of

$$\phi(R) = \frac{M_{bg}}{(R + a)} + \frac{M_{bh}}{R},$$

the ratio of tidal forces to globular cluster self-gravity is approximately given by:

$$\alpha = \frac{M_{bg}}{M_{gc}} \left( \frac{R_{gc}}{R + a} \right)^3 + \frac{M_{bh}}{M_{gc}} \left( \frac{R_{gc}}{R} \right)^3 .$$

The tidal radius is defined as the value of $R$ for which $\alpha \approx 1$. It can be seen from this expression that, for small values of $a$ the more massive spheroid component will determine the tidal radius. This formula can roughly predict the radius within which we expect significant globular cluster disruption, the black hole or spheroid parameters needed for disruption, or the radius at which these two effects are equal.

It is of interest to see how closely the single dimensionless parameter, $\alpha$, determines the mass stripped from a globular cluster. The mass loss from a globular cluster, given by our N-body simulation, is plotted vs. $\alpha$ in Fig. 7 for our various choices of parameters. Although there is a clear correlation, i.e. larger $\alpha$ result in larger mass loss, there is a variation between models. In particular, cases with a smaller black hole mass and a smaller $a$ can have a smaller $\alpha$ at $r_p$, but lead to the same overall mass loss. The analytic estimate of $\alpha$ at the point of closest approach in the orbit is not an exact indication of the mass loss because the cluster is subject to disruption over a larger part of its orbit. Cases with smaller $a$ and $M_{bh}$ have an $\alpha$ that declines less rapidly with increasing distance from pericenter, thus will be subject to more disruption for a larger period of time.

This analytic estimate has been extended in order to calculate the mass loss from a globular cluster as a function of distance from galactic center, $R$. For every $R$ there is a specific radius in the globular cluster for which $\alpha = 1$ and the mass outside this radius in the King model distribution can be taken as an estimate of stripped mass. This estimate was found to agree, to within a few percent, with our N-body simulations for cases with small mass loss. Deviations became substantial for cases with small $r_p$ (larger mass loss) because in these cases much of the disruption occurred at larger $R$.

For simplicity, we have considered only globular clusters with a particular value of concentration, expressed in terms of a half-mass radius. Clearly, smaller values of the half mass radius (for the same cluster mass) will lead to less disruption. It was found, in fact, that a cluster with a half mass radius of 2pc is not significantly disrupted outside a galactocentric distance of 50pc, even for the most massive black hole and most concentrated spheroid that we considered. The process of tidal disruption, both for black holes and spheroids, will only be effective for clusters that are relatively loose. Since the black hole masses were expressed in terms of the cluster masses, it can be seen that a more massive cluster (with the same half mass radius) would only be disrupted by a black hole that is more massive by the same factor. This means that the lower mass clusters will be most affected by tidal disruption, in agreement with Aguilar, Hut, and Ostriker (1988).

Values of $a$ of 400pc and 800pc may be representative of the bulges of spiral galaxies. Our own galaxy has been modeled with a half-mass radius of $\sim 2000$pc, which corresponds closely to the $a = 800$pc case. The degree of concentration of the bulge, or more generally, the mass distribution within the central few kpc of a galaxy, are crucial in determining the amount of tidal disruption of globular clusters. This had been illustrated in a statistical sense by Aguilar, Hut, and Ostriker (1988) who compared the galaxy models of Caldwell and Ostriker (1983) and of Bahcall, Schmidt and Soneira (1983). The latter model, with an added central nucleus was found to give a larger disruptive effect. We find (in Fig. 3b) that, except for the extreme case of a $10^9 M_\odot$ black hole, the bulge parameters are likely to be more important than a central black hole in determining the evolution of the globular cluster distribution. This suggests it would be useful to explore the possibility that bulge properties along the spiral sequence Sa - Sc could result in a systematic difference in the globular cluster distributions in these types of galaxies. Only if the spheroid is not highly concentrated, and if the cluster passes very close to the center of the galaxy, can a black hole of $10^8 M_\odot$ or less substantially affect the amount of disruption.

As mentioned in the introduction, some elliptical galaxies are likely to have black holes even more massive than $10^9 M_\odot$ at their centers. These galaxies will have spheroids that are quite extended, and thus the black hole will be the more important effect in most ellipticals. Although the mass of the spheroid has increased, the corresponding increase in $a$ balances this effect. A $10^{12} M_\odot$ spheroid would need to have $a = 1600\mathrm{pc}$ in order for the disruptive effects to enhance those contributed by the black hole alone (to increase the tidal radius by a factor of $\sim 2$). If $a$ is increased to 3200pc, the spheroid again has negligible effect. Depending on the shapes of the initial orbital distribution of globular clusters, a considerable region at the center of a giant elliptical can be evacuated (Ostriker, Binney, & Saha 1989). This is consistent with observations of the relatively less concentrated distributions of globulars, as compared to the light distribution, in M87 (McLaughlin, Harris, & Hanes 1994). The distribution of globular clusters in ellipticals should thus be strongly dependent on the central black hole mass, while in spirals, the bulge concentration will be the most important factor.

We conclude that a concentrated spheroid and a supermassive black hole can have similar effects on the globular cluster distribution. The debris that is stripped from the cluster has a narrow eccentricity distribution peaked around the original eccentricity of the globular cluster orbit. Globular clusters that pass close to the center of a galaxy will be preferentially destroyed, and these will tend to be the ones with orbits close to radial. Modeling of the full population of globular clusters is needed to assess the total number of stripped stars, however, the high eccentricity orbits predicted by the mechanism of tidal disruption are qualitatively consistent with the orbital properties of the halo star population.

We thank Robin Ciardullo and David Chernoff for numerous discussions and helpful suggestions. P.L. is supported in part by the NASA (at Los Alamos National Laboratory), NSF Young Investigator award PHY-9357219, and NSF grant PHY-9309834.

| $a$ | $M_{bh}$ | $r_p$ | $m_{gc}$ | $m_b$ | $m_u$ |
|-----|----------|-------|----------|-------|-------|
| 800 | 0        | 50    | 69.2     | 12.0  | 18.8  |
| 800 | 0        | 100   | 80.2     | 7.7   | 12.1  |
| 800 | 0        | 200   | 89.9     | 4.3   | 5.8   |
| 800 | 0        | 400   | 95.7     | 1.7   | 2.6   |
| 800 | $10^7$   | 50    | 70.6     | 12.9  | 16.5  |
| 800 | $10^7$   | 100   | 81.3     | 7.9   | 10.8  |
| 800 | $10^7$   | 200   | 89.8     | 4.2   | 6.0   |
| 800 | $10^7$   | 400   | 95.6     | 2.0   | 2.4   |
| 800 | $10^8$   | 50    | 45.6     | 25.9  | 28.5  |
| 800 | $10^8$   | 100   | 74.2     | 10.7  | 15.1  |
| 800 | $10^8$   | 200   | 87.6     | 5.4   | 7.0   |
| 800 | $10^8$   | 400   | 95.3     | 1.7   | 3.0   |
| 800 | $10^9$   | 50    | 0.0      | 47.6  | 52.4  |
| 800 | $10^9$   | 100   | 14.8     | 40.6  | 44.6  |
| 800 | $10^9$   | 200   | 59.1     | 23.9  | 17.0  |
| 800 | $10^9$   | 400   | 84.3     | 8.3   | 7.4   |
| 400 | 0        | 50    | 32.6     | 32.8  | 34.6  |
| 400 | 0        | 100   | 49.7     | 23.4  | 26.9  |
| 400 | 0        | 200   | 66.0     | 17.4  | 16.6  |
| 400 | 0        | 400   | 79.5     | 11.2  | 9.3   |
| 400 | $10^7$   | 50    | 27.0     | 34.9  | 38.1  |
| 400 | $10^7$   | 100   | 47.9     | 25.3  | 26.8  |
| 400 | $10^7$   | 200   | 65.2     | 18.0  | 16.8  |
| 400 | $10^7$   | 400   | 79.3     | 10.3  | 10.4  |
| 400 | $10^8$   | 50    | 21.1     | 37.0  | 41.9  |
| 400 | $10^8$   | 100   | 43.2     | 26.5  | 30.3  |
| 400 | $10^8$   | 200   | 63.4     | 17.7  | 18.9  |
| 400 | $10^8$   | 400   | 78.7     | 10.2  | 11.1  |
| 400 | $10^9$   | 50    | 0.0      | 49.7  | 50.3  |
| 400 | $10^9$   | 100   | 17.3     | 39.9  | 42.8  |
| 400 | $10^9$   | 200   | 43.0     | 29.5  | 27.5  |
| 400 | $10^9$   | 400   | 64.3     | 18.4  | 17.3  |

Table 1: Summary of N-body Simulation Parameters and Results.

## Figure Captions

Figure 1: Fraction of the original mass of the globular cluster that remains bound is plotted as a function of the pericentric distance for parabolic orbits. The four curves, in order of increasing destruction, are for no black hole, $10^7 M_\odot$, $10^8 M_\odot$, and $10^9 M_\odot$ black holes. Top and bottom plots are for spheroids with $a = 800$pc and $400$pc respectively. The larger destruction results from the more concentrated spheroid.

Figure 2: Same as in Fig. 1 but showing for the two spheroid distributions the fraction of the original mass of the globular cluster that remains bound as a function of the black hole masses. The four curves, in order of decreasing destruction, are for pericentric distances 50pc, 100pc, 200pc, and 400pc, respectively. The black hole is seen to have a large destructive effect only for close passages to the center.

Figure 3: The effect of changing the degree of bulge concentration is illustrated relative to the destructive effect of supermassive black holes of various masses. a) The fraction that remains bound to the globular cluster is shown as a function of black hole mass for various spheroid concentrations. The case with no black hole is plotted on the horizontal scale at $10^6 M_\odot$ for convenience of display. The three curves, in order of decreasing destruction, are for softening parameters 400pc, 800pc and 1600pc. All the cases correspond to orbits with pericentric distance of 100pc. b) Here it is illustrated that the bulge softening parameter is the most important factor in determining the fraction of mass stripped from the globular cluster. The four curves, in order of greater destruction, are for no black hole, $10^7 M_\odot$, $10^8 M_\odot$, and $10^9 M_\odot$ mass black holes, respectively. The plot of the fraction retained as a function of $a$ shows little difference between the curves that represent different black hole masses, except for the $10^9 M_\odot$ case.

Figure 4: The energy, angular momentum, and eccentricity distributions are examined for the debris that has been stripped from the globular cluster. Two cases are considered for $a = 400$pc, one with no black hole and one with a black hole of $10^9 M_\odot$. a) The bound (negative energy) and unbound (positive energy) material is seen as two peaks in the distribution, with wider peaks in the case with a black hole. b) The angular momentum distributions are given for bound (solid histogram) and unbound (dotted histogram) debris. The arrow represents the initial angular momentum of the center of mass of the globular cluster. c) The eccentricity distribution is given for stripped material that remains bound to the spheroid/black hole system. This shows that tidal stripping can widen the distribution, but that the effect is small.

Figure 5: The eccentricity distribution of material that is stripped from the globular cluster, but still bound to the spheroid/black hole system is given, as in Fig. 4c, but for runs with eccentricity of 0.8. For the two cases illustrated (with black hole masses of 0 and $10^9 M_\odot$) the spheroid concentration is $a = 400$pc and the closest approach distance is $r_p = 100$pc.

Figure 6: The "string of pearls" effect for globular cluster debris is illustrated. The time sequence shown illustrates the appearance of debris as it makes five orbits of the galaxy's center.

Figure 7: The percentage of the globular cluster mass that is stripped in our N-body simulations is plotted against the dimensionless parameter $\alpha$. Each curve, representing a particular choice of $a$ and $M_{bh}$, connects points with four different $r_p$ values: 50, 100, 200, and 400pc going from larger to smaller values of $\alpha$. Each curves is labeled with the black hole mass. Solid curves are used to represent cases with bulge concentration $a = 400$pc and dotted curves represent cases with $a = 800$pc.

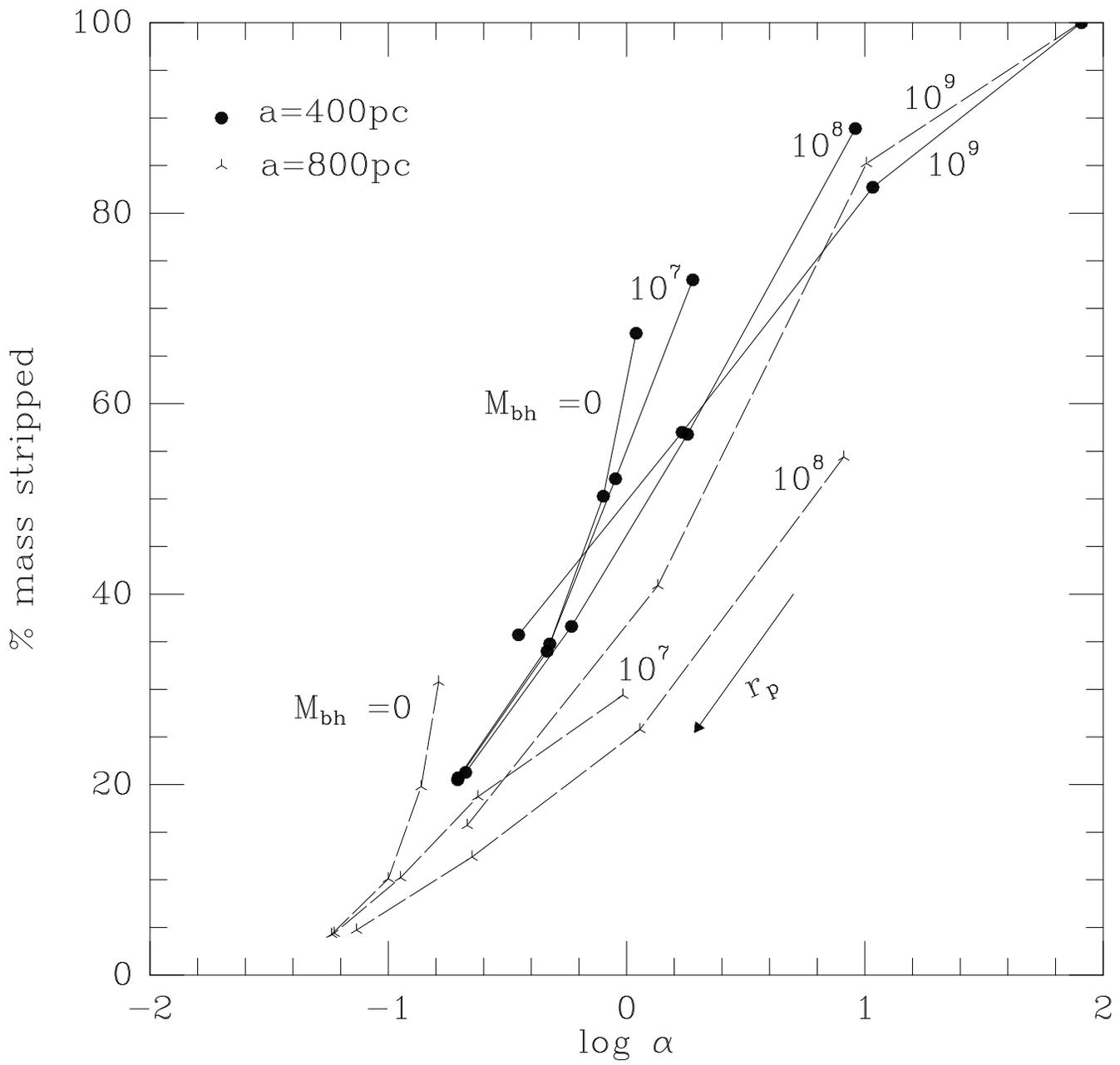